# Deterministic switching of perpendicular magnetization using Néel order-engineered out-of-plane spin in a single ferromagnet

Baiqing Jiang[1,2,\*], Ziqian Cui[1,2,\*], Hanying Zhang[1,2], Yuan Wang[1,2], and C. Bi[1,2,†]

[1]Laboratory of Microelectronics Devices and Integrated Technology, Institute of Microelectronics, Chinese Academy of Sciences, Beijing 100029, China

[2]University of Chinese Academy of Sciences, Beijing, 100049, China

[\*]These authors contributed equally to this work.

[†]clab@ime.ac.cn

**Abstract**

Perpendicular switching of a ferromagnet induced by spin torques is crucial for building high density spin-based memory and logic devices, where out-of-plane spin polarization ($\sigma_z$) has become a long sought-after goal for deterministic switching without assisted magnetic fields. Here we report the observation of $\sigma_z$ and resultant field-free perpendicular switching in a single ferromagnet without any spin torque generation layers, where $\sigma_z$ is achieved through the self-generated spin polarization in the ferromagnet that is engineered by the Néel order of an adjacent antiferromagnetic insulator. We further demonstrated that $\sigma_z$ emerges when the self-generated spin polarization is collinear with the Néel vector, where the spin current is reflected back to the ferromagnet, along with rotated spin polarization toward the out-of-plane direction to induce $\sigma_z$. Since no current is shunted by antiferromagnetic insulators and the Néel order does not rely on single-crystalline materials, these results may provide a CMOS-compatible solution for constructing energy-efficient field-free spintronic devices.

**I. Introduction**

Spin-transfer torque induced by a reference layer in magnetic sandwiched structures and spin-orbit torque (SOT) generated by SOT-materials are two fundamental spin torques. Particularly, SOT has emerged as an efficient means recently to manipulate the magnetized states of ferromagnetic functional layers (FFLs) in spintronic memory and logic devices [1–10], where SOT is usually generated through an additional SOT layer and the generated spin current penetrates into the adjacent FFL, inducing magnetization switching [1–3], domain nucleation [4–6], domain wall motion [7,8], and magnetization precession [9] for information storage or processing. In most SOT materials, the spin current occurs because of the spin Hall or Rashba effects [1–3], in which a z-directional spin current ($\mathbf{J_s}$) induced by an x-directional charge current ($\mathbf{J_c}$) can only have the spin polarization ($\mathbf{\sigma}$) along the y direction, limited by $\mathbf{\sigma} \propto \mathbf{J_c} \times \mathbf{J_s}$. However, in practical spintronic devices, particularly magnetic random-access memory (MRAM), perpendicularly magnetized ferromagnetic layers (along the z direction) with high scalability and thermal stability [11] are adopted, where the y-directional $\mathbf{\sigma}$ cannot induce deterministic z-directional magnetization switching without an assisted magnetic field [2,3]. Therefore, out-of-plane spin polarization along the z direction ($\sigma_z$) promising field-free switching of the perpendicular magnetization is urgent for practical high-density spintronic devices. So far, $\sigma_z$ was only reported in limited ferromagnetic multilayers with single-crystalline SOT layers, in which single-crystalline topological [12–14], antiferromagnetic [15–17], and low-dimensional materials [18–22] serve as the additional SOT layer and $\sigma_z$ strongly depends on the crystal orientations. In addition to the well-known technique difficulties for growing such single crystals on silicon substrates using complementary metal-oxide-semiconductor (CMOS) compatible technologies, it is very challenging for precisely controlling the preferred crystalline direction and thickness of the SOT layers (several nm, depending on the spin diffusion length) to reduce total write energy consumption by balancing SOT generation efficiency and current shunting in the device fabrication level.

Recently, $\mathbf{\sigma}$ along the y direction ($\sigma_y$) generated in ferromagnets (FMs) through the counterparts of anomalous Hall effects, anisotropic magnetoresistance, or planar Hall effects have been proposed [23–28]. Unfortunately, $\mathbf{\sigma}$ always cancels out in a single FM due to the bulk inversion symmetry since the directions of generated spin in the top and bottom sides are opposite [26,28], as shown in Fig. 1(a). As a result, the total $\mathbf{\sigma}$ as well as induced spin torque ($\tau_{tot}$) is zero, which make the self-generated spin current useless in practical devices for switching the FM itself unless used like conventional SOT materials, as the additional SOT layer, to rotate another FFL [24]. Moreover, when a FM severs as the SOT layer, a space layer is required to eliminate direct exchange coupling between the FM and FFL, which inevitably degrades the spin transmission toward FFLs and reduces spin efficiencies. The typical example utilizing FM as the SOT layer is NiFe/Ti/CoFeB trilayer, where **σ** generated from the NiFe (SOT generation layer) passes through Ti (space layer) and induces magnetization switching in CoFeB (FFL) [24]. Therefore, breaking the bulk inversion symmetry of a FM to induce the spin imbalance within the FM may provide another manner for spin generation without the additional reference or SOT layers, which is also crucial for utilizing the self-generated spin current.

Here we suggest that an adjacent antiferromagnetic insulator (AFI), whose Néel vector (**n**) interacts with the spin current at one side of a FM, may break the bulk inversion symmetry of the FM and induce non-zero **σ**. Specifically, in an AFI/FM bilayer, when **n** of AFI is collinear to the FM-generated **σ** at the AFI/FM interfaces, the self-generated spin current is reflected at the interface [29–31] and converges into spin backflow (**J$_b$**) that acts like SOT generated in conventional SOT material/FM bilayers, as shown in Fig. 1(b). We consider an AFI/FM structure because AFI does not shunt current, not cause extra energy consumption in application, and can also exclude the SOT originated from AFM itself [15–17] to make the spin torque generation mechanism clear. Moreover, **σ** may precess along **n** at the FM/antiferromagnet interfaces [32,33] and induce $\sigma_z$ in the spin backflow. In contrary, when **n** is perpendicular to **σ**, the self-generated spin current is absorbed at the interface, in which the bulk inversion symmetry is also broken and a pure spin torque may also form in the FM itself, but no $\sigma_z$ is expected as shown in Fig. 1(c). Note that the spin reflection and absorption may not be necessary to maintain unusual magnetoresistance [34]. So far, the non-zero **σ** and possible $\sigma_z$ in a single FM without additional SOT layers have not been achieved. In this work, we use the typical MRAM material, CoFeB, as the FM and NiO$_x$ as the AFI to observe non-zero **σ** as well as $\sigma_z$ in a single FM in the absence of additional SOT layers. We demonstrate that $\sigma_z$ can be gained in NiO$_x$/CoFeB bilayers, which is large enough to switch the perpendicularly magnetized CoFeB layer itself in a field-free mode. Since there is no single-crystalline material and conducting SOT layer involved, the demonstrated field-free $\sigma_z$ switching may be compatible with CMOS and present MRAM technologies with high energy efficiencies.

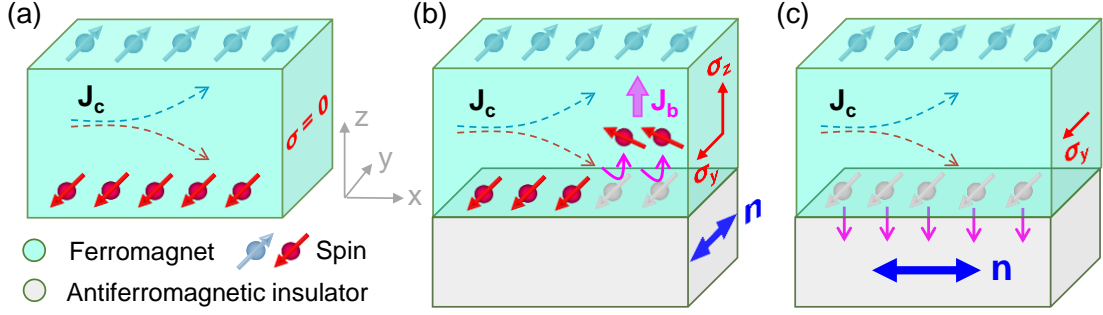

FIG. 1. Schematic of spin torque generation in a single ferromagnetic layer. (a) No pure **σ** and total spin torque generated in a single ferromagnet due to bulk inversion symmetry, where opposite spin formed at top and bottom sides always cancels out. **J**$_c$ is the charge current. (b) When an adjacent AFI whose **n** is collinear to spin at the AFI/FM interface presents, the self-generated spin in FM is reflected by the interface, along with rotated **σ** toward the z direction, resulting in a spin backflow, **J**$_b$, with both σ$_y$ and σ$_z$. Therefore, non-zero **σ**, particularly, σ$_z$ is formed. (c) When **n** is perpendicular to spin at the AFI/FM interface, the self-generated spin near the AFI/FM interface is absorbed by the AFI, where the balance of σ$_y$ between top and bottom sides is broken and non-zero σ$_y$, but not σ$_z$, is formed.

## II. Characterization of σ$_z$ in AFI/FM bilayers

The AFI/FM bilayers with a full structure of Sub/NiO$_x$ 30 nm/CoFeB 10 nm/SiO$_2$ 3 nm were deposited by using magnetron sputtering, where Sub represents Si/SiO$_2$ 300 nm substrates and the SiO$_2$ is a capping layer. The structures of Sub/MgO 10 nm/CoFeB 10 nm/SiO$_2$ 3 nm without antiferromagnets were also sputtered as the control samples. As shown in Fig. 2(a), the deposited films were patterned into 100 μm × 10 μm strips, sat in an Au coplanar waveguide (CPW), for characterizing spin torques through the spin-torque ferromagnetic resonance (ST-FMR) measurements [35]. The Néel vector, **n**, of the NiO$_x$ layer is aligned to be parallel (**I**$_{RF}$ // **n**) or perpendicular (**I**$_{RF}$ ⊥ **n**) to the microwave current direction through field-cooling operations (see Appendix for experimental methods). The self-generated **σ** in FM also follow **σ** ∝ **J**$_c$ × **J**$_s$ [26,28], and thus, **I**$_{RF}$ ⊥ **n** and **I**$_{RF}$ // **n** correspond to the configurations of Fig. 1(b) and Fig. 1(c), respectively. We first confirm that there is no σ$_z$ detected in the MgO/CoFeB control samples. Figure 2(b) shows the ST-FMR spectra of MgO/CoFeB samples when the applied external field (H) reverses, where the detected ST-FMR signals, V$_{mix}$, can overlap each other when the angle between H and applied microwave current (I$_{rf}$), φ, changes 180°, by considering a sign reversal. This is the typical ST-FMR behavior of conventional SOT materials where only σ$_y$ presents [18]. Generally, there should be no any ST-FMR signal detected in a single FM due to bulk inversion symmetry, but several works have reported that weak SOT may be induced in a single FM because of asymmetric interface scattering or nonuniform element distributions [36,37], similar to the Fig. 2(b). Sharply contrary to Fig. 2(b), ST-FMR spectra of NiO$_x$/CoFeB samples become asymmetric at positive and negative H and V$_{mix}$ and -V$_{mix}$ cannot overlap when H reverses, indicating the possible $\sigma_z$ generation [15,16,18]. The amplitude of ST-FMR peaks is also about five times larger than that of MgO/CoFeB samples. In particular, when **I$_{RF}$** $\perp$ **n** corresponding to the schematic of Fig. 1(b), the difference between V$_{mix}$ and -V$_{mix}$ (Fig. 2(c)) is more pronounced, compared to that of **I$_{RF}$** // **n** (Fig. 2(d)).

To quantitatively analyze **σ**, the ST-FMR spectra are further deconvoluted into a symmetric and an antisymmetric Lorentzian peak by using

$$V_{mix} = V_s \frac{\Delta H^2}{\Delta H^2 + (H-H_0)^2} + V_a \frac{(H-H_0)\Delta H}{\Delta H^2 + (H-H_0)^2}, \tag{1}$$

where V$_s$ and V$_a$ are the two coefficients describing the contributions of symmetric and antisymmetric Lorentzian peaks respectively, H$_0$ is the resonant field, and ΔH is the linewidth. According to the principle of ST-FMR measurements, V$_s$ originates from in-plane spin torques that include $\sigma_y$-induced damping-like torque and $\sigma_z$-induced field-like torque, while V$_a$ originates from out-of-plane spin torques that include $\sigma_y$-induced field-like torque and $\sigma_z$-induced damping-like torque [15,16]. Here the possible $\sigma_x$-induced spin torques are ignored because the angle-dependent ST-FMR results do not support the presence of $\sigma_x$. To further identify $\sigma_z$ contribution, extractions of V$_s$ and V$_a$ at positive and negative H are required, where V$_s$ showing the same sign at positive and negative H can be considered as a clear signal of the presence of $\sigma_z$ since $\sigma_y$-contributed V$_s$ and V$_a$ always show opposite signs at positive and negative H. The deconvoluted symmetric and antisymmetric Lorentzian peaks for MgO/CoFeB and NiO$_x$/CoFeB samples are shown in Fig. 2(b), Fig. 2(e) (**I$_{RF}$** $\perp$ **n**), and Fig. 2(f) (**I$_{RF}$** // **n**), respectively. For the MgO/CoFeB samples, V$_s$ (V$_a$) shows the same value but with opposite signs at positive and negative H (Fig. 2(b)), consistent with the presence of only $\sigma_y$. In contrary, Fig. 2(e) shows that the symmetric peaks point to the same direction when **I$_{RF}$** $\perp$ **n**, indicating the same sign of V$_s$ and providing substantial evidences of $\sigma_z$ existence [15]. According to the spin swapping mechanisms, when the in-plane polarized spin ($\sigma_y$) is reflected at the ferromagnetic interface, it can also generate a secondary spin current with out-of-plane spin polarization ($\sigma_z$) due to the spin precession under an interfacial spin-orbit field [33], as observed in Fig. 2(e). When **I$_{RF}$** // **n**, the extracted V$_s$ (V$_a$) show different values and opposite signs, indicating the presence of σ_y and weak σ_z.

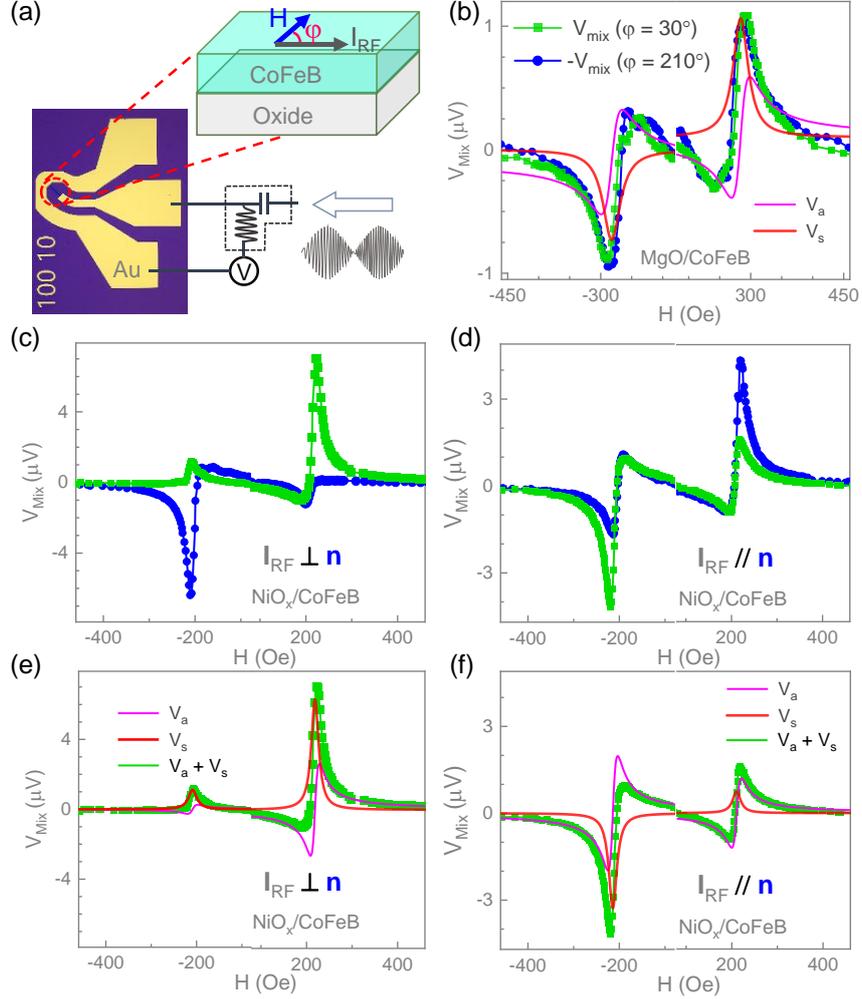

FIG. 2. Characteristics of $\sigma_z$ in ST-FMR spectra. (a) Schematic of ST-FMR configurations and optical images of CPW devices. (b) Acquired ST-FMR spectra of MgO/CoFeB devices at φ = 30° and 210°. (c), (d) Comparison of ST-FMR spectra at φ = 30° and 210° when (c) **I_RF** ⊥ **n** and (d) **I_RF** // **n** in NiO_x/CoFeB devices. (e), (f) Deconvolution of ST-FMR spectra at φ = 30° using symmetric and antisymmetric Lorentzian functions when (c) **I_RF** ⊥ **n** and (d) **I_RF** // **n** in NiO_x/CoFeB devices. The solid lines in (b), (e), and (f) are the deconvoluted symmetric and antisymmetric peaks. The frequency and power of microwave are 5 GHz and 15 dBm, respectively.

The angle-dependent ST-FMR measurements can further be used to evaluate $\sigma_y$ and $\sigma_z$ contributions [15,16,18]. Figure 3(a-c) show the extracted $V_s$ and $V_a$ as a function of φ for NiO_x/CoFeB with **I_RF** ⊥ **n** (Fig. 3(a)) and **I_RF** // **n** (Fig. 3(b)) and MgO/CoFeB (Fig. 3(c)) samples. $V_s$ contains the $\sigma_y$-induced damping-like torque ($S_{DL}^y$) and $\sigma_z$-induced field-like torque contributions ($S_{FL}^z$), which follows

$$V_s = S_{DL}^y \cos\varphi \sin 2\varphi + S_{FL}^z \sin 2\varphi; \qquad (2)$$

$V_a$ contains the $\sigma_y$-induced field-like torque ($A_{FL}^y$; the Oersted field contribution is also included in this term) and $\sigma_z$-induced damping-like torque contributions ($A_{DL}^z$), which follows

$$V_a = A_{FL}^y \cos\varphi \sin 2\varphi + A_{DL}^z \sin 2\varphi. \qquad (3)$$

The solid lines in Fig. 3(a-c) show the contributions of each torque and fitting results using Eq. (2) and Eq. (3). First, in all three cases, both $V_s$ and $V_a$ experimental data can be well fitted using Eq. (2) and Eq. (3) even though the $V_s$ signal of MgO/CoFeB samples is very weak, indicating that the $\sigma_x$-induced torques can be ignored. Second, for the NiO$_x$/CoFeB samples, both the $\sigma_z$-induced damping-like and field-like contributions when **I$_{RF}$** $\perp$ **n** are much stronger than those of **I$_{RF}$** // **n**, confirming that strong $\sigma_z$ is generated in the **I$_{RF}$** $\perp$ **n** case, as illustrated in Fig. 1(b). For the MgO/CoFeB samples, both $V_s$ and $V_a$ data can be well fitted by considering the $\sigma_y$ and the Oersted field contributions only, agreeing with previous observations that pure $\sigma_y$ may be induced in a single FM due to asymmetric interface scattering or nonuniform element distributions. Third, by comparing the three cases, we can conclude that $\sigma_z$ does generate in a single FM and can be modulated by controlling the Néel vector of adjacent antiferromagnets. Moreover, $\sigma_y$ can also be induced when **I$_{RF}$** // **n** due to the symmetry breaking induced by spin absorption at the AFI/FM interface. All these results are consistent with the spin generation mechanisms shown in Fig. 1.

To estimate the relative contribution of $\sigma_z$ when **I$_{RF}$** $\perp$ **n** and **I$_{RF}$** // **n**, we further calibrate the $\sigma_y$-induced field-like and Oersted field contributions in both cases using the field-compensated measurements [38] (see Appendix A for experimental methods). As schematically shown in inset of Fig. 3(e), by using a Hall bar structure, a small y-directional external field, $h_y$, is applied simultaneously when a current (**I**) is applied along the x direction. Due to the current-induced field-like torques, the in-plane magnetization oscillates slightly under alternate +I and -I, resulting in a change of planar Hall signal ($\Delta V_{yx}$). As shown in Fig. 3(d), when $h_y = 0$ Oe, $\Delta V_{yx}$ depends on magnetization directions, which can clearly reveal the magnetization switching induced by an x-directional field ($H_x$). However, when the applied $h_y$ cancels the effective field ($H_{FL}^y$) of current-induced field-like torques, the in-plane magnetization does not oscillate and $\Delta V_{yx}$ disappears [38]. Figure 3(d) shows the recorded $\Delta V_{yx}$ as a function of $H_x$ when I = ± 2.0 mA in both **I** $\perp$ **n** and **I** // **n** cases. When $h_y = 0.2$ Oe, $\Delta V_{yx}$ signals disappears around $H_x = 0$ in both cases, indicating that $H_{FL}^y$ is cancelled by h$_y$, and thus $H_{FL}^{y}$ = -0.2 Oe. When h$_y$ = 0.5 Oe, ΔV$_{yx}$ reverses because the total y-directional field, $H_{FL}^{y}$ + h$_y$, change signs, validating the reliability of this measurement. Figure 3(e) shows that the measured $H_{FL}^{y}$ is proportional to the applied current and equals when **I** ⊥ **n** and **I** // **n**. Since the σ$_y$-induced field-like torque and the Oersted field, corresponding to $A_{FL}^{y}$, are the same in both **I** ⊥ **n** and **I** // **n** cases, we can use $A_{FL}^{y}$ as the reference to estimate each type of spin torques. Figure 3(f) summarizes the extracted $S_{FL}^{z}/A_{FL}^{y}$, $A_{DL}^{z}/A_{FL}^{y}$, and $S_{DL}^{y}/A_{FL}^{y}$ ratios from Fig. 3(a) and 3(b), where the much larger σ$_z$ contribution can be observed when **I$_{RF}$** ⊥ **n** ($A_{DL}^{z}/A_{FL}^{y}$: 0.44 when **I$_{RF}$** ⊥ **n**; 0.049 when **I$_{RF}$** // **n**). Moreover, a sizeable σ$_y$ can also be induced when **I$_{RF}$** // **n**. Note that the finite $A_{DL}^{z}/A_{FL}^{y}$ when **I$_{RF}$** // **n** that is one order smaller than that in the **I$_{RF}$** ⊥ **n** configuration can be attributed to a slight misalignment between **I$_{RF}$** and **n** during **n** manipulation (see Appendix A for experimental methods).

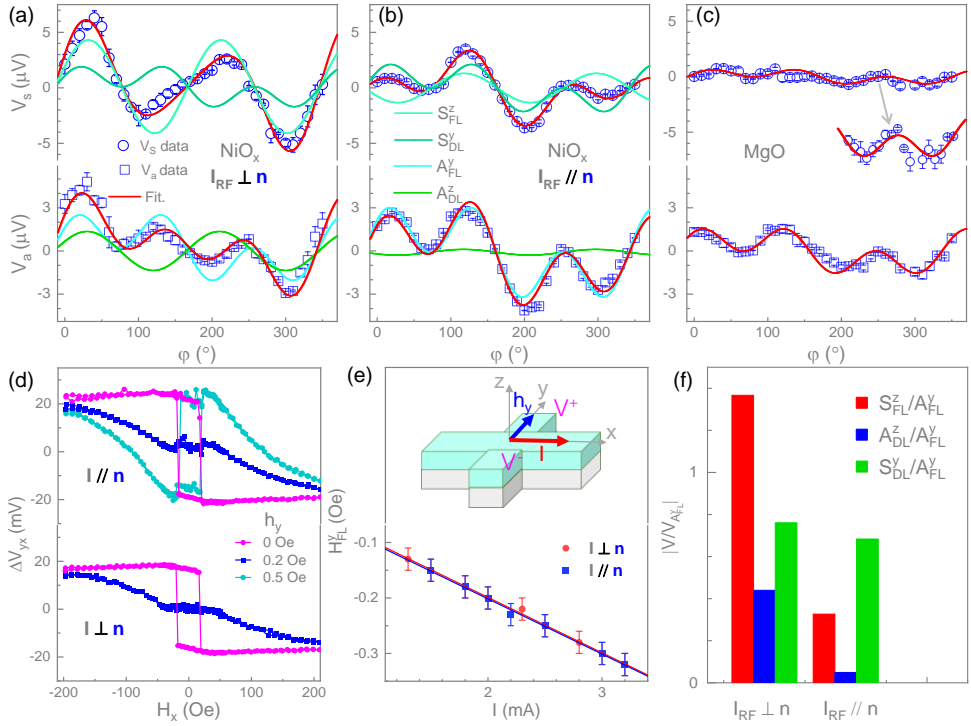

FIG. 3. Quantitative analysis of in-plane and out-of-plane spin torques. (a-c) The extracted symmetric and antisymmetric parts of ST-FMR as a function of φ for the NiO$_x$/CoFeB devices when (a) **I$_{RF}$** ⊥ **n** and (b) **I$_{RF}$** // **n** and (c) MgO/CoFeB devices. The solid lines are the fitting results and separated various spin torque contributions using Eq. (2) and Eq. (3). (d) Representative ΔV$_{yx}$ as a function of H$_x$ measured under different h$_y$ for the NiO$_x$/CoFeB devices during field-compensated measurements. (e) The determined $H_{FL}^{y}$ of NiO$_x$/CoFeB devices through the recorded ΔV$_{yx}$ versus H$_x$ curves as a function of applied current. Inset shows the schematic of field-compensated measurements. (f) The calculated $S_{FL}^{z}/A_{FL}^{y}$, $A_{DL}^{z}/A_{FL}^{y}$, and $S_{DL}^{y}/A_{FL}^{y}$ ratios when **I$_{RF}$** ⊥ **n** and **I$_{RF}$** // **n**.

**III. $\sigma_z$-induced deterministic field-free switching**

We further verify whether $\sigma_z$ is large enough to drive deterministic perpendicular magnetization switching in the absence of assisted in-plane magnetic fields. As schematically shown in Fig. 4(a), the full structure for demonstrating field-free perpendicular switching is Sub/CoFeB 5 nm/NiO$_x$ 30 nm/Ti 0.8 nm/CoFeB 1.2 nm/MgO 2 nm/TaO$_x$ 2 nm. The bottom 5 nm CoFeB shows in-plane magnetic anisotropy, which directly couples to the NiO$_x$ layer and can be used to manipulate the Néel vector of the NiO$_x$ layer through direct coupling field during field-cooling operations (see Appendix A for experimental methods). Similar to previous reports, the 0.8 nm Ti layer is used for enhancing perpendicular magnetic anisotropy (PMA) of the top 1.2 nm CoFeB layer, in which the spin torques from the 0.8 nm Ti layer itself can be ignored and meanwhile the 0.8 nm thickness can still allow spin transmission across the Ti layer [24]. The 2 nm MgO layer is used for building interfacial PMA [39] and TaO$_x$ is the capping layer. Moreover, the 30 nm NiO$_x$ can also eliminate the direct coupling between bottom 5 nm and top 1.2 nm CoFeB layers.

We first confirm that PMA can be established by measuring the anomalous Hall effects. Figure 4(b) shows the Hall resistance (R$_H$) as a function of perpendicular field (H$_z$) when **I** $\perp$ **n** and **I** // **n**, where the sharp H$_z$-driven switching and square R$_H$ loop indicate strong PMA established in both cases. To estimate the possible $\sigma_z$-induced effective perpendicular field ($H_{eff}^z$) that is particularly important to drive deterministic magnetization switching, a current pulse (I$_p$) is applied before recording R$_H$ at each H$_z$ step during R$_H$ versus H$_z$ measurements. The shift of R$_H$ versus H$_z$ loops reflects I$_p$-induced $H_{eff}^z$ [40]. As shown in Fig. 4(c), when **I** $\perp$ **n**, opposite shift of R$_H$ loops can be observed clearly when I$_p$ = ± 10 mA, consistent with the opposite $H_{eff}^z$ induced by +I$_p$ and -I$_p$. The Joule heating can be excluded as the origin of R$_H$ loop shift because it does not depend on current direction. When **I** // **n**, no opposite shift of R$_H$ loops can be observed even with a larger I$_p$ = ± 15 mA, as shown in Fig. 4(d). Therefore, these results have also demonstrated the presence of $\sigma_z$ when **I** $\perp$ **n** that results in a $H_{eff}^z$, consistent with the ST-FMR measurements.

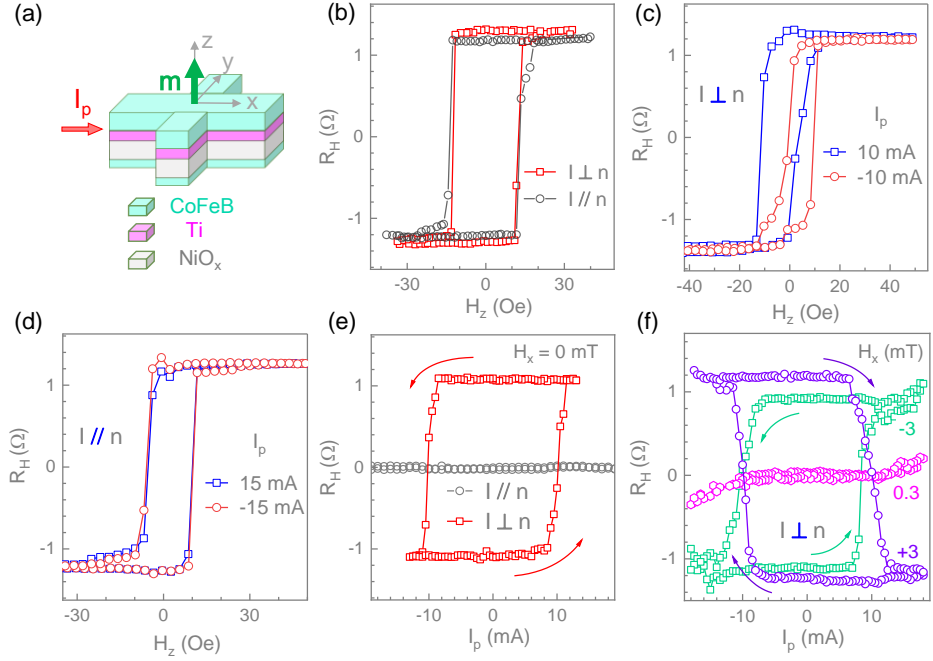

FIG. 4. Demonstration of $\sigma_z$-induced field-free perpendicular switching. (a) Schematic of perpendicularly magnetized $NiO_x$/CoFeB devices and experimental configurations. (b) Recorded $R_H$ as a function of $H_z$ when $\mathbf{I} \perp \mathbf{n}$ and $\mathbf{I} // \mathbf{n}$ using a 25 μA dc detection current. (c), (d) Recorded $R_H$ after $I_p$ as a function of $H_z$ when (c) $\mathbf{I} \perp \mathbf{n}$ and (d) $\mathbf{I} // \mathbf{n}$. (e) $R_H$ as a function of applied $I_p$ measured without any applied external magnetic fields demonstrates current-driven field-free switching when $\mathbf{I} \perp \mathbf{n}$. No switching behavior appears when $\mathbf{I} // \mathbf{n}$. (f) Current-driven magnetization switching loops under different $H_x$ when $\mathbf{I} \perp \mathbf{n}$.

Figure 4(e) shows $R_H$ as a function of $I_p$ in the absence of any external fields, where the field-free switching can be achieved when $\mathbf{I} \perp \mathbf{n}$. The critical switching current is about ± 10 mA for the 5 μm Hall bars, corresponding to the critical switching current density of 32.26 MA/cm$^2$ by considering uniform current distribution in both bottom and top CoFeB layers. For comparison, when $\mathbf{I} // \mathbf{n}$, no current-driven magnetization switching is observed, highlighting the role of $\sigma_z$ in the field-free switching. The current-driven perpendicular switching under a positive or negative in-plane field, $H_x$, has also been observed when $\mathbf{I} \perp \mathbf{n}$, as shown in Fig. 4(f), where the current-driven switching directions depend on the direction of $H_x$, similar to conventional SOT-driven perpendicular switching [1–3]. The $R_H$ versus $I_p$ loop is clockwise for $H_x$ = +3 mT, while it becomes anticlockwise for $H_x$ = -3 mT. When $H_x$ = 0.3 mT, no switching behaviors are observed, indicating that the $\sigma_z$-determined field-free anticlockwise switching (Fig. 4(e)) is fully compensated by $H_x$ = 0.3 mT induced clockwise switching, and thus, there is no deterministic magnetization switching. The observation of current-driven perpendicular switching not only directly evidences the presence of non-zero **σ** in a single FM engineered by controlling **n** of the adjacent AFI, but also demonstrates current-driven field-free perpendicular switching can be achieved using the self-generated spin torque without additional SOT layers. It should be mentioned that $\sigma_z$ can also be generated at a NiO$_x$/Pt interface, where the polarization originates from the Pt layer and the field-free switching of an adjacent ferromagnet has also been demonstrated [41].

**IV. Technology discussion**

SOT-driven perpendicular switching has been proposed to replace classic spin-transfer torque (STT) in MRAM with high speed, energy efficiency, and reliability because no incubation time is required in the initial switching processes [42] and the large write current does not need to pass through the MgO tunnel layer [2]. However, it suffers from an external magnetic field during write operations, which cannot be integrated in high-density MRAM cells [43]. In principle, the inversion symmetry has to be broken to realize deterministic switching if only $\sigma_y$ exists in conventional SOT materials. Lateral symmetry breaking like a nonuniform oxide layer or nonuniform PMA in the film plane [40] does not suit for large-scale fabrication. Out-of-plane symmetry breaking by creating composition gradient along the thickness direction is usually achieved in a thick ferromagnet, corresponding a large critical switching current [44,45], and cannot be realized in a magnetic tunnel junction (MTJ) structure with large tunnel magnetoresistance (TMR) ratio for read operation. The build-in in-plane magnetic fields such as exchange bias fields by introducing additional magnets [46,47] may result in large writing errors besides the MTJ-compatible problems. Recently reported $\sigma_z$ in single-crystalline materials [13–20,22,48] also encounters the same MTJ-compatible problems like deposition of CoFeB layers (the only functional ferromagnetic materials in MRAM) with PMA, large TMR, and growth of single-crystal in large Si wafers. Moreover, control of the thickness of SOT layers and etching processes with atomic precision is also very challenging for nanofabrication [49]. The demonstrated $\sigma_z$-switching in this work is based on the perpendicularly magnetized CoFeB layers and does not rely on single-crystalline materials, and therefore, there are no the large-scale grown and MTJ-compatible problems. The critical switching current density also approaches that based on topological materials promising high charge-spin conversion efficiencies [12]. Since there is no current shunting in the AFI layer, the thickness of AFI can also be adjusted to facilitate nanofabrication processes.

**V. Conclusion**

We have utilized the self-generated spin torque to switch a perpendicularly magnetized FM in the absence of external fields and SOT layers, sharply contrary to present SOT technologies where an assisted in-plane field and SOT generation layer are the two basic requirements for the current-driven perpendicular magnetization switching. We achieve this by using an AFI to modulate the spin scattering and absorption at the AFI/FM interfaces, which breaks the inversion symmetry of the FM and results in a non-zero $\sigma$ as the source of spin torque generation. ST-FMR measurements have been used to characterize the self-generated spin torques, where $\sigma_y$ can be detected in both $\mathbf{I} \perp \mathbf{n}$ and $\mathbf{I} // \mathbf{n}$ cases, while $\sigma_z$ can only be observed when $\mathbf{I} \perp \mathbf{n}$. The observed $\sigma_z$ is responsible for the current-driven field-free perpendicular switching when $\mathbf{I} \perp \mathbf{n}$. Compared to conventional SOT switching and other field-free solutions, the demonstrated self-generated spin torque switching does not require additional SOT layer and single-crystalline materials, and thus, is suitable for large-scale fabrication using CMOS-compatible technologies.

**Acknowledgements**

This work was supported by the Beijing Natural Science Foundation under Grant No. Z230006, the MOST of China under Grant 2019YFB2005800, the Natural Science Foundation of China under Grant 61974160, and the Strategic Priority Research Program of the Chinese Academy of Sciences under Grant No. XDB44000000.

**Appendix A: Experimental methods**

**1. Sample preparation**

All samples were deposited using magnetron sputtering with the base vacuum better than $1 \times 10^{-8}$ Torr before sputtering. The CoFeB, Ta, and Ti layers were sputtered from their metal targets through dc sputtering under 2 mTorr Ar pressure, in which the 2 nm Ta layer naturally oxidizes to $TaO_x$ as the capping layer for the perpendicularly magnetized samples. $NiO_x$, MgO, and $SiO_2$ layers were deposited through rf sputtering under 0.6 mTorr Ar pressure, where ~ 2% $O_2$ was also flowed when sputtering $NiO_x$. For the ST-FMR measurements, the $NiO_x$/CoFeB samples were first patterned into strips with the width of 10 μm and then CPW structures (Ti 5 nm/Au 100 nm) were deposited on the strips, where the dimension of $NiO_x$/CoFeB layers sat in the center line of CPW is 10 μm × 100 μm. For the field-free switching measurements, the samples were patterned into a Hall bar structure with the width of 5 μm (along the current direction) and 2 μm bars for Hall voltage detection. To manipulate the Néel vector of NiOx layers, a field-cooling process was performed, where the fabricated ST-FMR or Hall bar devices were first heated to 300 °C for 30 s and then cooled down to the room temperature under a 0.6 T external field. The external field was set perpendicular or parallel to the current direction during field-cooling processes to realize **I** ⊥ **n** or **I** // **n** configurations. The perpendicularly magnetized samples were also annealed at 300 °C for additional 5 mins to enhance PMA.

**2. ST-FMR measurements**

For ST-FMR measurements, an amplitude-modulated microwave current, $I_{rf}$, with the frequency between 2 GHz and 12 GHz and power of 15 dBm was applied to the $NiO_x$/CoFeB or MgO/CoFeB devices using CPW structures. The microwave-generated rectified signal was separated by a bias-tee and then picked up by a lock-in amplifier. The angle between applied in-plane magnetic field (H) and $I_{rf}$ is defined as φ, which can be tuned by rotating H, as shown in Fig. 2(a). The ST-FMR spectra were acquired at a fixed frequency and φ by sweeping H at room temperature. The ST-FMR spectra were further deconvoluted into a symmetric and an antisymmetric Lorentzian peak, where $V_s$ and $V_a$ were extracted at each frequency and φ.

**3. Electrical measurements**

For the field-compensated measurements to estimate field-like torques and the Oersted field, in addition to an external field, $H_x$, applied to switch magnetization, a small y-directional field, $h_y$, was also applied. At each $H_x$ step, the Hall voltage ($V_{yx}$) was recorded at a positive current and a positive $h_y$ first, and then $V_{yx}$ under a negative current and a negative $h_y$ was also recorded. $\Delta V_{yx} = V_{yx}(+I, +h_y) - V_{yx}(-I, -h_y)$ was then calculated and plotted as a function of $H_x$ like Fig. 3(d). By changing the strength and direction of $h_y$, the current-induced y-directional effective field, $H_{FL}^y = -h_y$, can be determined when $\Delta V_{yx}$ versus $H_x$ does not show any loops. For the current-driven magnetization switching (Fig. 4(e) and 4(f)) and $H_{eff}^z$ measurements (Fig. 4(c) and 4(d)), a current pulse, $I_p$, was applied first, and then, a 25 μA dc current was applied to measure $R_H$. Figure 4(c) and 4(d) show the measured $R_H$ as a function of $H_z$, while Fig. 4(e) and 4(f) show the measured $R_H$ as a function of $I_p$.

**Appendix B: Verification of Néel vectors**

The Néel vector, **n**, after field-cooling operations was verified through anisotropic magnetoresistance (AMR) measurements, from which the in-plane magnetization switching field can be determined at positive and negative field directions. If **n** is manipulated to collinear with current, an exchange bias field ($H_{ex}$) should be induced and can be determined through the positive and negative switching field. Figure 5 shows the resistance as a function of in-plane field applied along the current direction when **I** ⊥ **n** and **I** // **n** configurations defined through the field-cooling operations. As expected, $H_{ex}$ was only observed when **I** // **n**, indicating that **n** can be manipulated by field-cooling operations.

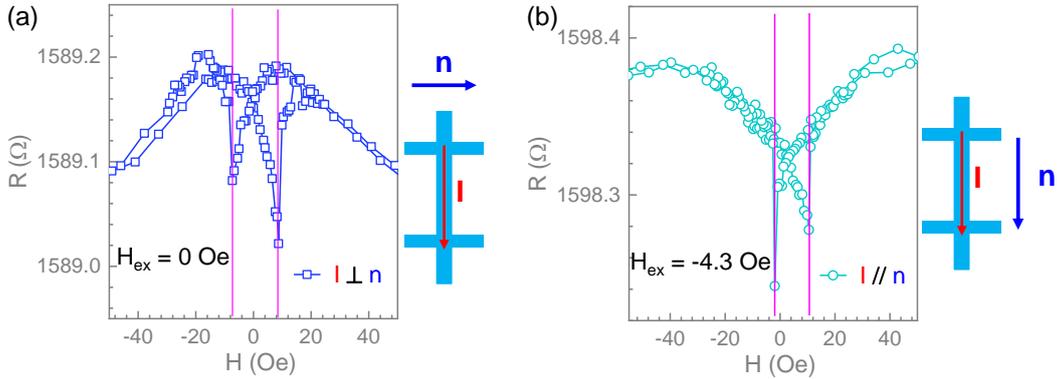

FIG. 5. R as a function of in-plane magnetic field when (a) **I** ⊥ **n** and (b) **I** // **n** for the $NiO_x$ 30 nm/CoFeB 10 nm Hall bar devices. The dimension of Hall bars is 10 μm × 100 μm.